\documentclass[11pt]{article}
\usepackage{amscd,amsmath,amsthm,amsfonts,amssymb}
\usepackage{graphicx}
\usepackage{multicol}
\usepackage{setspace}%\doublespacing
\numberwithin{equation}{section}

\newcommand{\comment}[1]{}

\newcommand{\Real}{\mathrm{Re}}
\newcommand{\Imag}{\mathrm{Im}}

\def\sech{{\rm sech}}
\def\tanh{{\rm tanh}}

\def\Ai{{\rm Ai}}
\def\Bi{{\rm Bi}}

\def\uinf{u_{\infty}}

\def\AllInt{ \displaystyle\int_{-\infty}^{\infty}}
\def\BlackInt{\displaystyle\int_{-\uinf z}^{\uinf z}}
\def\MInt{\displaystyle\int_{S_L(\zeta)}^{S_R(\zeta)}}

\title{\bf\Large Perturbations of Dark Solitons}

\author{Mark J. Ablowitz$^1$, Sean D. Nixon$^1$,Theodoros P. Horikis$^2$,
and Dimitri J. Frantzeskakis$^3$}

\date{\normalsize \today }
\begin{document}
\maketitle
\begin{abstract}

A method for approximating dark soliton solutions of the nonlinear Schr\"{o}dinger equation under the influence of perturbations is presented.  The problem is broken into an inner region, where core of the soliton resides, and an outer region, which evolves independently of the soliton.  It is shown that a shelf develops around the soliton which propagates with speed determined by the background intensity.  Integral relations obtained from the conservation laws of the nonlinear Schr\"{o}dinger  equation are used to approximate the shape of the shelf.  The analysis is developed for both constant and slowly evolving backgrounds.  A number of  problems are investigated including linear and nonlinear damping type perturbations.

 \end{abstract}

Perturbation theory as applied to solitons which decay at infinity, i.e. so-called bright solitons, has been developed over many years  cf. \cite{Karpman1977, Kodama1981, HermanRL1990}.
The analytical work employs a diverse set of methods including perturbations of the inverse scattering transform (IST), direct multi-scale perturbation analysis, perturbations of conserved quantities etc; the analysis applies to a wide range of physical problems.  In optics, a central equation which describes the envelope of a quasi-monochromatic wave-train  is the nonlinear Schr\"{o}dinger (NLS) equation  \ref{NLS}, which in normalized form is given by
\begin{equation*}
iU_{z} + \frac{D}{2} U_{tt} + n|U|^2 U =0
\label{NLS}
\end{equation*}
where $D,n$ are constant.  In this paper we will consider the NLS equation in a typical nonlinear optics context where $D$ corresponds to the group-velocity dispersion (GVD), $n>0$  is related to the the nonlinear index of refraction, $z$ is the direction of propagation and $t$ corresponds to the retarded time.  In this form, the sign of $D$ determines whether the light focuses or defocuses.  In the anomalous GVD (or self-focusing nonlinearity) regime the NLS eq. exhibits so-called "bright" solitons which are pulses which decay rapidly at infinity. In this case the solitons are formed due to a balance between dispersion and self-focusing  cubic nonlinearity. 

On the other hand, in the normal GVD (or self-defocusing nonlinearity) regime decaying pulses broaden and bright solitons of the NLS eq. do not exist.  Instead solitons can be found as localized dips in intensity which decay off of a continuous-wave (cw) background.  These dark solitons, which are termed black when the intensity of the dip goes to zero and grey otherwise, are also associated with a rapid change in phase across the pulse.  The experimental observations of dark solitons in both fiber optics \cite{Emplit1987} and planar waveguides \cite{Andersen1991} sparked significant interest in the asymptotic analysis of their propagation dating back two decades.

Using the above mentioned analytical methods, the propagation of bright solitons under perturbation is described by the adiabatic evolution of the soliton parameters; i.e. the soliton's height, velocity, position shift and phase shift.  However, the non-vanishing boundary of dark solitons introduces serious complications when applying the perturbative methods developed for bright solitons.  In early work, the particular case of added linear loss was studied both numerically \cite{Zhao1989} and analytically \cite{Giannini1990}.  The analysis was specifically for black solitons and solved explicitly for higher order correction terms.  These results were re-derived \cite{Lisak1991} with a more straightforward method. 
The method  was extended to grey solitons and general perturbations but only for the two of the four main soliton parameters; background height and soliton depth, were determined.  The evolution of the background was shown to be independent of the soliton by Kivshar \cite{Kivshar1994} where the asymptotic behavior at infinity was used to separate the propagation of the background magnitude from the rest of the soliton.  The amplitude/width of the soliton `core' was determined via a perturbed Hamiltonian.   Of the methods proposed many have employed perturbation theory based on IST theory. In  \cite{Konotop1994} orthogonality conditions are derived from a complete set of squared Jost functions (eigenfunctions of the linearized NLS operator) \cite{Kaup1976}; from these conditions the soliton paramters are, in principle, determined; over the years various corrections/modifications  have been made to the details \cite{Chen1998, Lashkin2004, Ao2005}.  

In this paper we address a central issue systemic through all these methods.  For dark solitons finding the adiabatic evolution of the soliton parameters (background height, soliton depth, position shift and phase shift) alone is insufficient to fully characterize the evolution of a dark soliton.  We find both analytically and numerically the existence of a shelf which develops around a dark soliton under perturbation.  The tendency for shelves to generate around dark solitons under external perturbation \cite{Burtsev97} was used to explain discrepancies in the perturbed conservation laws.  But, the analytical calculation of the core soliton parameters was not obtained. Subsequently the shelf contribution has been  ignored.  However, the shelf is critical in developing the perturbation theory and has a non-trivial contribution to the integrals employed to find expressions for the soliton parameters.  In this paper, we use perturbed conservation laws since they can be easily derived directly from the NLS equation and do not require the associated subtleties inherent in the IST method and extends to non-integrable problems.  To carry out the procedure we employ suitable asymptotic information about the higher order perturbation terms (beyond the soliton); in general we do not need the exact higher order solution to solve the leading order problem for the key parameters.  Finally we note that shelves in soliton perturbation theory have been found earlier in a different class of problems. They were needed to effectively understand the KdV equation under perturbation %cf. \cite{AND REVIEW IN MJA and SEGUR + KNICKERBOCKER AND NEWELL 1980}.
 In the KdV eq there is a small shelf produced in the wake of the soliton. The height/speed of the soliton, shelf and the additional soliton parameter which determines the center of the soliton are all determined by perturbation theory \cite{Kodama1981}.**

The outline of this paper is as follows.  In Section \ref{Sec: Boundary} we pose the problem and illustrate how the background evolves under perturbation independent of any localized solitary wave disturbances.  Sections \ref{Sec: First Correction} - \ref{Sec: ConLawsBlack} set up the basic analysis and a prototypical problem is discussed which helps describe  the ideas.  The method of multiple scales is employed to find the the first order approximation for a black soliton under the action of a dissipative perturbation which decays to zero well away from the soliton core.  The concept of a moving boundary layer is used to bridge the differences between the inner soliton solution and the outer background.  This discrepancy between the approximate soliton solution and the background manifests itself as a shelf developing on either side of the soliton.  Perturbed conservation laws are used to find the growth of the shelf in both magnitude and phase.  The analytic results are shown to be in agreement with numerical simulations of the perturbed NLS equation.  In Sections \ref{Sec: General} - \ref{Sec: GenCon}  the method is extended to grey solitons under general perturbations.  Suitable asymptotic information about the shelf is obtained from the linear first order perturbation equation; the complete solution of the linear problem is not required.  In Sections \ref{Sec: Dissipation} - \ref{Sec: TPA} the perturbation method is applied to some physically relevant perturbations: dissipation and two photon absorption.  We find that the spatial frequency of the soliton differs from that of the background that it resides on.  All of the adiabatically varying core soliton  parameters and the shelf have not been obtained in the many previous studies of perturbed dark solitons.  In Appendix \ref{Adx: Sec} we derive secularity conditions from Fredholm alternative type arguments which agree with the results found from the perturbed conservation laws.

%FOR COMPLETENESS: IN APPENDIX ADD 
%1) THEODOROS CHANGE OF VARIABLES TO NONEVEOLING BACKGROUND. ADD ENOUGH TO SHOW THAT IT GIVE SAME RESULTS
%2) ADD EXAMPLES: JUST MAIN RESULTS
%A) F[U]= I GAMMA EPSILON U
%B)DISPERSIVE PERTURBATION --FROM THEODOROS BEC PROBLEM; ADD RESULTS TO SHOW THAT THEY YIELD A SHELF

\section{The Boundary at Infinity}
\label{Sec: Boundary}

Let us consider the problem of the NLS eq. with normal dispersion $D= -1$, $n = 1$ and with an additional small forcing perturbation (we can always rescale NLS to get these unit values)
\begin{equation}
iU_{z} - \frac{1}{2} U_{tt} + |U|^2 U = \epsilon F[U]
\label{NLSPert}
\end{equation}
where $|\epsilon| \ll 1$.  Further we will assume a non-vanishing boundary value at infinity; i.e., $|U| \not \rightarrow 0$ as $t \rightarrow \pm \infty$. The effect the perturbation has on the behavior of the solution at infinity is independent of any local phenomena such as pulses which do not decay at infinity; i.e. dark solitons.  In the case of a a continuous wave background, which is relevant to perturbation problems with dark solitons as well as in applications to lasers, we have $U_{tt} \rightarrow 0$ as $t \rightarrow \pm \infty$ and the evolution of the background at either end $U \rightarrow  U^{\pm}(z)$ is given by the equation
\begin{equation}
\label{BigUInf}
i\frac{d}{dz} U^{\pm} + | U^{\pm}|^2  U^{\pm} = \epsilon F[ U^{\pm}]
\end{equation}
which we can break up into a magnitude and phase $ U^{\pm}(z) = u^{\pm} (z) e^{i\phi^{\pm}(z)}$ where $u^{\pm}(z)>0$ and $\phi^{\pm}(z)$ are both real functions of $z$
\begin{subequations}
\begin{eqnarray}
\frac{d}{dz} u^{\pm} &=& \epsilon \Imag \left[ F[u^{\pm} e^{i \phi^{\pm}} ] e^{-i \phi^{\pm}} \right] \\
\frac{d}{dz} \phi^{\pm} &=&  (u^{\pm})^2 - \epsilon\Real \left[ F[u^{\pm} e^{i \phi^{\pm}} ] e^{-i \phi^{\pm}}\right]/u^{\pm} 
\end{eqnarray}
\label{Background1}
\end{subequations}
The above equations completely describe the adiabatic evolution of the background under the influence of the perturbation $F[u]$.  Although this is true for all choices of perturbation, we will further restrict ourselves to perturbations which maintain the phase symmetry of equation \eqref{NLSPert}; i.e., $F[U(z,t) e^{i\theta} ] = F[U(z,t)] e^{i\theta}$.  This is a sufficient condition to keep the magnitude of the background equal on either side and a property of most commonly considered perturbations.  We assume that at $z=0, u^+(0)=u^-(0),$ then, since  $u^{\pm}(z)$ satisfy the same eq., the evolution is the same for all z.  Hence $u^+(z)=u^-(z) \equiv \uinf(z)$.  While this restriction is convenient the essentials of the method presented here apply in general. The equations for the background evolution \eqref{Background1} can now be further reduced by considering the phase difference $\Delta \phi_{\infty}(z) = \phi_{\infty}^+(z) -\phi_{\infty}^-(z) $ which is the parameter related to the depth of a dark soliton (see below); here $ \phi^{\pm}(z)$ represents the phase as $t \rightarrow \pm \infty$ respectively
\begin{subequations}
\begin{eqnarray}
\frac{d}{dz} \uinf &=& \epsilon \Imag \left[ F[\uinf  ] \right] \\
\frac{d}{dz} \Delta\phi_{\infty} &=&  0
\end{eqnarray}
\label{Background2}
\end{subequations}
Thus, while the magnitude of the background evolves adiabatically the phase difference remains unaffected by the perturbation.  

Let us now focus on the evolution of a dark soliton under perturbation.  To simplify our calculations we take out the fast evolution of the background phase 
\begin{equation}
U = u e^{\int_0^z \uinf(s)^2 ds}
\label{U_to_u_trans}
\end{equation}
 so equation \eqref{NLSPert} becomes
\begin{equation}
iu_z - \frac{1}{2} u_{tt} + (|u|^2 - \uinf^2) u = \epsilon F[u]
\label{NormalNLS}
\end{equation}
The dark soliton solution to the unperturbed equation is given by
\begin{equation}
u_s(t,z) = \left( A + i B \tanh \left[ B \left( t -A z- t_0 \right) \right] \right) e^{i\sigma_0}
\label{darksoliton1}
\end{equation}
where the core parameters of the soliton: $A,B,t_0,\sigma_0$ are all real, the magnitude of the background is $(A^2+B^2)^{1/2}= \uinf$ and the phase difference across the soliton is $2\tan^{-1}\left( \frac{B}{A} \right)$, $A \neq 0$.  When $A=0$ equation \eqref{darksoliton1} describes a black soliton, which has a phase difference of $\pi$.
%%%%%%%%%
%%%%%%%Still Needs Work

Below, we employ the method of multiple scales by introducing a slow scale variable $Z = \epsilon z$ with the parameters $A$, $B$, $t_0$ and $\sigma_0$ being functions of $Z$.  A perturbation series solution for equation \eqref{NormalNLS} is assumed
\begin{equation}
u= u_0(Z,z,t) + \epsilon u_1(Z,z,t) + O(\epsilon^2)
\end{equation}
The first order approximation $u_0(Z,z,t)$  should satisfy the slowly varying boundaries from equations \eqref{Background2}, which means two of the parameters are already pinned down $A(Z) = \uinf(Z) \cos \frac{\Delta\phi_{\infty}}{2} $ and $ B(Z) = \uinf(Z) \sin  \frac{\Delta\phi_{\infty}}{2}$ where we identify $\sigma_0= \frac{  \phi^{+}+ \phi^{-}}{2}$..  This, however, means that the shape of the soliton at first order is determined by non-local effects.  This, however, means that the shape of the soliton at first order is determined by non-local effects and this suggests that higher order terms might be required in order to characterize how the soliton evolves under perturbation.
%%%%%%%%%%%%
%%%%%%%%%%%%%

\section{The First Order Correction}
\label{Sec: First Correction}

%
%Solving for $u_1(Z,z,t)$ explicitly may be difficult and is not ideal for a general method.  Nevertheless, for clarity we will begin with an example in which higher order terms can be found.  This will also provide useful insight for the general problem
We write the solution in terms of the amplitude and phase: $u = q e^{i \phi}$ where $q$ and $\phi$ are both real functions of $z$ and $t$ so equation \eqref{NormalNLS} becomes
%%
%\begin{equation}
%F[u] = i \gamma u_{tt}, ~~\gamma>0
%\end{equation}
%%
%and at a leading order we consider a black pulse.  We write the solution in terms of the amplitude and phase: $u = q e^{i \phi}$ where $q$ and $\phi$ are both real functions of $z$ and $t$ so equation \eqref{NormalNLS} becomes
%
\begin{equation*}
iq_z - \phi_z q - \frac{1}{2} \left( q_{tt} + i2\phi_t q_t + q(i \phi_{tt} - \phi_t^2) \right) +  (|q|^2 - \uinf^2) q = \epsilon F[u]
\end{equation*}
Once we introduce the additional multiple scale variable $Z= \epsilon z$; the real and imaginary parts of the above eq. are:
\begin{subequations}
\begin{eqnarray*}
q_z &=& \frac{1}{2} (2\phi_t q_t + q \phi_{tt}) + \epsilon \left(\Imag\left[F[u] \right] - q_Z\right) \\
\phi_z q &=& -\frac{1}{2}(q_{tt} - \phi_t^2 q) + (|q|^2 - \uinf^2) q + \epsilon \left( \Real\left[F[u] \right] - \phi_Zq \right)
\end{eqnarray*}
\end{subequations}
Expanding $q$ and $\phi$ as series in $\epsilon$: $q  = q_0 + \epsilon q_1 + O(\epsilon^2)$ and $\phi = \phi_0 +\epsilon \phi_1 + O(\epsilon^2)$, we have at  $O(1)$
\begin{subequations}
\begin{eqnarray}
q_{0z} &=& \frac{1}{2} (2\phi_{0t} q_{0t} + q_0 \phi_{0tt}) \\
\phi_{0z} q_0 &=& -\frac{1}{2}(q_{0tt} - \phi_{0t}^2 q_0) + (|q_0|^2 - \uinf^2) q_0
\end{eqnarray}
\end{subequations}
with the general dark soliton solution
\begin{subequations}
\begin{eqnarray}
q_0 &=& \left( A(Z)^2 + B(Z)^2 \tanh^2 (x) \right) ^{1/2} \\
\phi_0 &=& \tan^{-1} \left[ \frac{B(Z)}{A(Z)}\tanh (x) \right] + \sigma_0(Z)\\
 x &=& B \left( t - \int_0^z A(\epsilon s) ds  - t_0(Z) \right)
\end{eqnarray}
\label{Fullqphi}
\end{subequations}
%
%Since $F[u] = i\gamma u_{tt} \rightarrow 0$ as $t\rightarrow \pm \infty$,  the equations \eqref{Background2} reduce to $\frac{d}{dz} \uinf = \frac{d}{dz} \phi_{\infty} = 0$ and as a corollary $A(Z)$ and $B(Z)$ are both constants.  Furthermore, for a black soliton initial conditions $A(0)$ must be $0$ and $A(Z) = 0$, $B(Z) = \uinf$.  The form of the first order solution \eqref{Fullqphi} reduces to
For a black soliton the form of solution \eqref{Fullqphi} is taken to be
\begin{subequations}
\begin{eqnarray}
q_0(Z,z,t) &=& \uinf \tanh \left[  \uinf \left( t  - t_0(Z) \right)\right] \\
\phi_0(Z,z,t) &=& \sigma_0(Z)
\end{eqnarray}
\label{Blackqphi}
\end{subequations}
where we note that in this representation $q_0$ is allowed to be negative.

At $O(\epsilon)$ we have
\begin{subequations}
\begin{eqnarray}
q_{1z} &=& \frac{1}{2}\left[ 2(\phi_{0t}q_{1t} + q_{0t}\phi_{1t}) + q_0\phi_{1tt} + q_1\phi_{0tt}\right]  + \Imag[F] - q_{0Z}\\
\phi_{1z}q_0 &=& - q_1\phi_{0z} - \frac{1}{2}\left[ q_{1tt} - (2\phi_{0t}\phi_{1t})q_0 - \phi_{0t}^2 q_1 \right] + 3q_0^2q_1 - \uinf^2q_1 + \Real[F] - \phi_{0Z}q_0
\end{eqnarray}
\label{order1E}
\end{subequations}

Solving for $q_1, \phi_1$ may be difficult and is not ideal for a general method seeking to obtain the basic core parameters of the soliton to require finding its solution.  Nevertheless, for clarity we will begin with an example in which higher order terms can be found explicitly.  This will also provide useful insight for the general problem. Consider
\begin{equation}
F[u] = i \gamma u_{tt}, ~~\gamma>0
\label{utt}
\end{equation}
and for concreteness, at a leading order we assume a black pulse which then yields
\begin{subequations}
\begin{eqnarray}
q_{0Z} &=& - t_{0Z} q_{0t}\\
\phi_{0Z} &=& \sigma_{0Z}
\end{eqnarray}
\label{SlowZderivatves}
\end{subequations}
If we look for a stationary solution, $q_{1z}=\phi_{1z}=0$, and note that $\phi_{0t} = \phi_{0tt} = 0$ then equations \eqref{order1E} reduce to
\begin{subequations}
\begin{eqnarray}
0&=& \frac{1}{2} \left[ 2 (q_{0t}\phi_{1t}) + q_0 \phi_{1tt} \right] + \Imag[F] + t_{0Z}q_{0t}
\label{blackphase}\\
0 &=& - \frac{1}{2}q_{1tt} + 3q_0^2q_1 - \uinf^2q_1 + \Real[F] - \sigma_{0Z}q_0
\label{blackheight}
\end{eqnarray}
\end{subequations}
where $\Imag[F] = \gamma q_{0tt}$ and $\Real[F] = 0$.  
First we look at equation \eqref{blackphase} 
\begin{equation}
q_{0t}\phi_{1t} + \frac{1}{2}q_0\phi_{1tt}  = - \gamma q_{0tt} - t_{0Z} q_{0t}
\end{equation}
which after multiplying by $q_0$, using properties of the leading order solution and integrating, yields
\begin{equation} 
\phi_{1t} =  \frac{4}{3} \gamma q_0 - t_{0Z} + c_1q_0^{-2}
\end{equation}
Since $q_0^{-2}$ has a singularity at $t_0$ we set $c_1=0$ to have a bounded solution.  After integrating we are left with
 \begin{equation}
 \phi_1 = \frac{4}{3} \gamma \ln \left[ \cosh (\uinf (t-t_0))\right] - t_{0Z} t +c_2
\end{equation}
Asymptotically for large $t$ we have 
\begin{equation}
\phi_{1t}^+ = \frac{4}{3} \gamma \uinf - t_{0Z} , \ \ \ \ \phi_{1t}^- = -\frac{4}{3}\gamma \uinf - t_{0Z}
\label{FirstPAsym}
\end{equation}
where the superscript $^{\pm}$ indicates the value of a function as $t\rightarrow \pm \infty$ respectively.

We can also solve explicitly for $q_1$.  After a change of variables $x = \uinf (t - t_0)$ and substituting in for $q_0$ equation \eqref{blackheight} becomes
\begin{equation}
q_{1xx} + (6\tanh^2(x) - 4)q_1 = -2 \frac{\sigma_{0Z}}{\uinf} \tanh (x)
\label{BH2}
\end{equation}
The homogenous problem is now a special form of 
\begin{equation}
Q_{xx} + (n(n+1) \sech^2 x - n^2) Q = 0 
\end{equation}
which has the bounded solution
\begin{equation} 
Q = sech^n x 
\end{equation}
With this solution we can use reduction of order to solve equation \eqref{BH2} 
\begin{equation} 
q_1 = \left[ c_1 + c_2 \left( \frac{1}{4}\sinh(4x) + 2 \sinh (2x) + 3x \right) + \frac{\sigma_{0Z}}{8\uinf} \left(x - \frac{1}{4} \sinh (4x) \right) \right] \sech^2 (x)
\end{equation}
By looking at the asymptotic behavior at $x \rightarrow \pm \infty$ 
\begin{equation}
q_1 \sim \frac{1}{16} \left(c_2 - \frac{\sigma_{0Z}}{\uinf} \right) e^{\pm2x} 
\end{equation}
we see that to avoid blow up we must take $c_2 = \frac{\sigma_{0Z}}{\uinf}$.  Furthermore, we require that the full solution $q_1$ vanish at $t= t_0$ so that the $u$ remains anti-symmetric.  Now the unique solution to \eqref{blackheight} is 
\begin{equation} 
q_1 = \frac{\sigma_{0Z}}{4\uinf} \left[ \sinh(2 \uinf (t-t_0)) + 2\uinf (t-t_0) \right] \sech^2 (\uinf (t-t_0))
\end{equation}
Looking at the asymptotic behavior as $t \rightarrow \pm \infty$ we have
\begin{equation}
q_1 \rightarrow (2 \uinf^2)^{-1} \sigma_{0Z} q_0^{\pm} = \pm \frac{\sigma_{0Z}}{2\uinf}
\label{FirstAsym}
\end{equation}

\section{Boundary Layer}
\label{Boundary Layer}

Notice that $q_1 \not\rightarrow 0$ and $\phi_1 \not\rightarrow 0 $ as $t\rightarrow \pm \infty$.  As a result, the solution $u\approx (q_0+\epsilon q_1)e^{i(\phi_0 +\epsilon \phi_1)}$ to order $\epsilon$ does not match the boundary conditions at infinity.  Thus, our problem is now broken into two areas: the region which matches behavior at infinity and is unaffected by the soliton, and the region in which the $O(\epsilon)$ correction term is correct.   We introduce a boundary layer in which there is a transition from a nonzero value in the perturbation term to zero (see also \cite{Kodama1981,% Newell Knickerbocker
}).  Note in this section we will consider the more general case when $\uinf$ is a function of $Z=\epsilon z$.  We find the behavior of this boundary layer, where the regions are matched, by looking for a near constant wave solution.  For this we return to equation \eqref{NormalNLS} and seek a solution perturbed around the solution at infinity, say $u \approx (\uinf + \epsilon w) e^{i (\phi^{\pm} + \epsilon \theta)}$ where $w$ and $\theta$ are real functions of $z$ and $t$; the equation is satisfied at $O(1)$ and we have at $O(\epsilon)$
\begin{equation}
-\theta_z \uinf + iw_z - \frac{1}{2}\left[ i \uinf \theta_{tt} + w_{tt} \right] + 2\uinf^2 w = F[\uinf + \epsilon w] - \left( i\frac{d\uinf}{dZ} - \uinf \frac{d\phi^{\pm}}{dZ} \right)
\label{BLFull}
\end{equation}
After substituting in equations \eqref{Background1} and \eqref{Background2} the right hand side is $F[\uinf + \epsilon w] - F[\uinf] \approx \epsilon F[w]$.  Thus, the right hand side is actually a higher order term and may be dropped and as a corollary the boundary layer is independent of perturbation.  We now break \eqref{BLFull} into real and imaginary parts 
\begin{subequations}
\begin{eqnarray}
\theta_z \uinf &=&  2\uinf^2 w - \frac{1}{2} w_{tt}
\label{layer1a}\\
w_z &=& \frac{1}{2} \uinf \theta_{tt}
\label{layer1b}
\end{eqnarray}
\end{subequations}
Taking a derivative with respect to $z$ of  equation \eqref{layer1b} and then substituting in for $\theta_z$ we get
\begin{eqnarray*}
w_{zz} &=& \frac{1}{2} \uinf \theta_{ztt}\\
&=& \frac{1}{2} \uinf\left[ \frac{1}{\uinf} (2 \uinf^2 w - \frac{1}{2}w_{tt})_{tt} \right]\\
&=& \uinf^2 w_{tt} - \frac{1}{4}w_{tttt}
\end{eqnarray*}
Similarly, taking a derivative with respect to $z$ of equation \eqref{layer1a} and then substituting in for $w_z$ we get
\begin{eqnarray*}
\uinf \theta_{zz} &=& 2\uinf^2 w_z - \frac{1}{2} w_{ztt}\\
&=& 2\uinf^2(\frac{1}{2}\uinf \theta_{tt} ) - \frac{1}{2} (\frac{1}{2}\uinf \theta_{tt})_{tt}\\
&=& \uinf \left[ \uinf^2 \theta_{tt} - \frac{1}{4} \theta_{tttt}\right]
\end{eqnarray*}
Leaving us with
\begin{subequations}
\begin{eqnarray}
w_{zz} &=& \uinf^2 w_{tt} - \frac{1}{4}w_{tttt}
\label{layer2a}\\
\theta_{zz} &=& \uinf^2 \theta_{tt} - \frac{1}{4}\theta_{tttt}
\label{layer2b}
\end{eqnarray}
\label{Layer2}
\end{subequations}
which is the same equation for both functions, though we will need a different solution for each.  This is because of the differing boundary conditions to correctly match the inner region to the outer region.  To the left of the soliton these boundary conditions are 
\begin{subequations}
\begin{eqnarray}
w(-\infty) = 0 & w(\infty) = q_1^-\\
\theta(-\infty) = 0 & \theta_t(\infty) = \phi_{1t}^-
\end{eqnarray}
\label{leftBC}
\end{subequations}
To the right of the soliton these boundary conditions are 
\begin{subequations}
\begin{eqnarray}
w(-\infty) = q_1^+ & w(\infty) = 0\\
\theta_t(-\infty) =  \phi_{1t}^+ & \theta(\infty) = 0
\end{eqnarray}
\label{rightBC}
\end{subequations}

If we let  $w= e^{i(kt + \int_0^z \omega(z,k)dz)}$, then the 'dispersion' relation for equation \eqref{layer2a} is found to be
\begin{equation}
\omega^2 = \uinf^2(Z) k^2 + \frac{1}{4}k^4
\end{equation}
For long waves ($k \ll 1$) we have roughly $\omega(z,k) \approx \pm \uinf(z) k$ or $w = e^{ik(t \pm \int_0^z \uinf (\epsilon z))}$.  Thus, we see that long wave solutions (i.e. $|k| \ll 1$) move with instantaneous velocity $V(z) = \pm \uinf(z)$.  This is also true for equation \eqref{layer2b}.

With this in mind, we look for solutions to equations \eqref{Layer2} in a moving frame of reference: $x = t - Vz$ and $\zeta = z$.
\begin{eqnarray*}
w_{\zeta\zeta} &=&  2Vw_{\zeta x} + (\uinf^2 - V^2)w_{xx} - \frac{1}{4}w_{xxxx}\\
\theta_{\zeta\zeta} &=&  2V\theta_{\zeta x} + (\uinf^2 - V^2)\theta_{xx} - \frac{1}{4}\theta_{xxxx}
\end{eqnarray*} 
And, for $V = \pm \uinf$ 
\begin{subequations}
\begin{eqnarray}
w_{\zeta \zeta } &=&  2Vw_{\zeta x}  - \frac{1}{4}w_{xxxx}\\
\theta_{\zeta \zeta } &=&  2V\theta_{\zeta x}  - \frac{1}{4}\theta_{xxxx}
\end{eqnarray} 
\label{Layer3}
\end{subequations}

We assume that derivatives with respect to $x$ are small; i.e. long waves.  There are several ways to balance the terms in equations \eqref{Layer3} (see Appendix A), the optimal one being 
\begin{equation*} 
\partial_{\zeta \zeta } \ll \partial_{\zeta x} \sim \partial_{xxxx} \ll1
\end{equation*}\
leaving us with
\begin{subequations}
\begin{eqnarray}
\label{layer3a}
0 &=& 2V w_{\zeta x} - \frac{1}{4} w_{xxxx}\\
\label{layer3b}
0 &=& 2V \theta_{\zeta x} - \frac{1}{4} \theta_{xxxx}
\end{eqnarray}
\end{subequations}

There are now two similarity solutions which we find to satisfy the boundary conditions \eqref{leftBC} and \eqref{rightBC} derived from matching the two regions.  First, by making the transformation $\theta_x = \tilde{f}\left(\tilde{\xi}\right)$ and $\tilde{\xi} = x/\zeta ^{1/3}$ in equation \eqref{layer3b} we get 
\begin{equation*}
0 = \frac{2}{3}V\tilde{\xi}\tilde{f}'  - \frac{1}{4}\tilde{f}'''
\end{equation*}
which can then be further reduced by the transformation $f = \tilde{f}'$ and $\xi = -2\left(\frac{V}{3}\right)^{1/3} \tilde{\xi}$ to get
\begin{equation}
0 = f'' - \xi f 
\label{Airy}
\end{equation}

Equation \eqref{Airy} is the well known Airy equation% \cite{ABORMOWITZ AND STEGEN} 
with general solution $f(\xi) = c_1 \Ai(\xi) + c_2 \Bi(\xi)$ where $\Ai(\xi)$ and $\Bi(\xi)$ are special functions defined in terms of infinite series or improper integrals.  Since, $\Bi(\xi)$ grows exponentially we take $c_2$ to be $0$.  For $V = -\uinf$ we are looking for a solution $\theta$ which goes to zero as $x \rightarrow -\infty$ and as a direct result $\tilde{f} = \theta_x \rightarrow 0$ as $\tilde{\xi}  = x/\zeta ^{1/3}\rightarrow -\infty$.  With this we can now unwrap the transformations made earlier. If we consider the boundary conditions on the left of the soliton
\begin{subequations}
\begin{equation}
\theta(\zeta ,x) = c_1 \displaystyle\int_{-\infty}^x  \displaystyle\int_{-\infty}^{a\tilde{x}/\zeta ^{1/3}} \Ai(s)ds \tilde{x}
\label{BLPhaseleft}
\end{equation} 
where $a =  -2\left(\frac{V}{3}\right)^{1/3}$ and $c_1 = \phi_{0t}^-$.  Note that the sign of $a$ depends on the sign on $V$.  In the same way on the right of the soliton we find that when $V = \uinf$ the solution is
\begin{equation}
\theta(\zeta , x) = c_2 \displaystyle\int_{\infty}^x  \displaystyle\int_{-\infty}^{a\tilde{x}/\zeta ^{1/3}} \Ai(s)ds d\tilde{x}
\label{BLPhaseright}
\end{equation}
\end{subequations}
This solution matches the boundary conditions for the phase $\theta$ with $c_4 = \phi_{0t}^+$.

To get the other solution we begin by factoring out a derivative with respect to $x$ in equation \eqref{layer3a}
\begin{eqnarray*}
0 &=& \left(2Vw_\zeta  - \frac{1}{4}w_{xxx}\right)_x\\
c_3 &=& 2Vw_\zeta  - \frac{1}{4}w_{xxx}
\end{eqnarray*}
For this to satisfy the zero boundary condition (on either side) it must be $c_3 = 0$ leaving us with
\begin{equation*}
0 = 2Vw_{\zeta}  - \frac{1}{4}w_{xxx}
\end{equation*}
Which under the same procedure used above has solution
\begin{equation}
w(x) = c_4  \displaystyle\int_{-\infty}^{a x/\zeta ^{1/3}} \Ai(s)ds
\label{BLMag}
\end{equation}
for both $V= -\uinf$ and $V= \uinf$.  An important point is that there are two boundary layers moving away from the soliton solution with speed $\uinf$ generating a shelf.  This shelf has a non-negligible contribution in the integrals employed in soliton perturbation be it in the form of secularity conditions or the conservation laws which we will be employing.

\section{Perturbed Conservation Laws}
\label{Sec: ConLaws}

We still need to solve for the slowly evolving parameters $\sigma_0(Z)$ and $t_0(Z)$ for the black soliton.  This will be done by deriving equations for the growth of the shelf from the perturbed conservations laws associated with perturbed NLS \eqref{NormalNLS}.  The shelf is described by the asymptotic parameters $q_1^{\pm}$ and $\phi_{1t}^{\pm}$, which are in turn expressed in terms of $\sigma_{0Z}$ and $t_{0Z}$.  In general we will use the Hamiltonian $H$, the energy $E$, the momentum $I$, and the center of energy $R$.
\begin{subequations}
\begin{eqnarray} 
H &=& \AllInt \left[ \frac{1}{2} \left| \frac{\partial u }{\partial t} \right|^2 + \frac{1}{2}(\uinf^2 - |u|^2)^2 \right] dt\\
E &=& \AllInt \left[ \uinf^2 - |u|^2 \right]dt\\
I  &=& \AllInt \Imag\left[ u u^*_t\right] dt\\
R &=& \AllInt t \left( \uinf^2 - |u|^2 \right)  dt
\end{eqnarray} 
\end{subequations}
Note that since the standard total energy ($E_{Total} = \int|u|^2 dt$) would be infinite, we define the energy of a dark pulse to be the difference of the total energy and the energy of a continuous wave of corresponding magnitude.  For unperturbed NLS the first three integrals are conserved quantities while the last can be written in term of the momentum, i.e.; $\frac{dR}{dz} = -I$.  Evolution equations for these integrals may be easily obtained from equations \eqref{NLSPert} and \eqref{Background2}
\begin{subequations}
\begin{eqnarray} 
\label{dHdz}
\frac{dH}{dz} &=& E \frac{d}{dz}\uinf^2+   2\epsilon \Real \AllInt F[u] u_{z}^* dt \\
\frac{dE}{dz} &=& 2\epsilon \Imag \AllInt F[\uinf] \uinf - F[u] u^*dt\\
\frac{dI}{dz} &=&  2 \epsilon  \Real \AllInt  F[u] u_t^* dt\\
\frac{dR}{dz} &=& - I  + 2\epsilon \Imag \AllInt t \left( F[\uinf]\uinf - F[u]u^* \right)dt
\end{eqnarray} 
\label{ConEvol}
\end{subequations}
For our example problem only the energy and the momentum equations will be used.

\section{Full Black Solution}
\label{Sec: ConLawsBlack}

We begin with the perturbed conservation of energy
\begin{equation}
\frac{d}{dz}\AllInt  \left[ \uinf^2 - |u|^2 \right]  dt = 2\epsilon \Imag \AllInt F[\uinf] \uinf - F[u] u^*dt
\label{ConEnergy}
\end{equation}
Substituting in $u= q e^{i\phi}$, $F[u] = i\gamma u_{tt}$, $T = t - t_0$  and taking the terms up to $O(\epsilon)$ we have 
\begin{subequations}
\begin{equation}
\frac{d}{dz} \AllInt \left[ q_0^2 - \uinf^2 + \epsilon 2 q_0 q_1  \right]dT =  2\epsilon   \AllInt  \gamma q_{0TT} q_0 dT
\label{CE1}
\end{equation}
At $O(1)$ equation \eqref{CE1} is satisfied: $\frac{d}{dz} \int \left[ q_0^2 -\uinf^2\right] dT = 0$; At $O(\epsilon)$ we have 
\begin{equation}
\frac{d}{dz} \BlackInt q_0 q_1dT = -\gamma \AllInt q_{0T}^2 dT
\label{CE2}
\end{equation}
This is an equation for the change in energy caused by the propagation of the shelf.  Notice that on the left hand side of equation \eqref{CE2} we are only integrating over $T \in [-\uinf z , \uinf z]$, the inner region around the soliton defined by the boundary layers found in the last section.  Since $q_0$ and $q_1$ are only functions of $T$, we can apply the fundamental theorem of calculus to arrive at 
\begin{equation}
\uinf \left[ q_1(\uinf z) q_0(\uinf z) + q_1(-\uinf z) q_0(-\uinf z) \right] = -\gamma \uinf^3\frac{4}{3}
\end{equation}
And, for large $z$ (although in practice $\uinf z$ only needs to be modestly larger than the the full-width-half-max), we take $q_0 \rightarrow \pm \uinf$ and $q_1 \rightarrow q_1^{\pm}$ leaving us with 
\begin{equation}
q_1^+ - q_1^- = -\frac{4}{3} \uinf \gamma\
\label{EnergyResult}
\end{equation} 
\end{subequations}
By substituting in the asymptotic approximation \eqref{FirstAsym} found early for $q_1^{\pm}$, we arrive at an expression for $\sigma_0$
\begin{equation}
\sigma_{0Z} = - \gamma \frac{4}{3} \uinf^2 
\end{equation}

Next, we consider the modified conservation of momentum  
\begin{equation}
\frac{d}{dz} \Imag \AllInt  u u_t^* dt = 2 \epsilon  \Real \AllInt  F[u] u_t^* dt
\label{ConMo}
\end{equation}
Again, we let $u= q e^{i\phi}$, $F[u] = i\gamma u_{tt}$, $T = t - t_0$ and use the perturbation expansion for $u$ up to $O(\epsilon)$ so that equation \eqref{ConMo} becomes
\begin{subequations}
%\begin{equation}
%\frac{d}{dz} \Imag \int q (q_t + i \phi_t q)dt = \epsilon 2 \Real \int i \gamma q_{0tt} q_{0t}^* dt 
%\end{equation}
%
\begin{equation}
-\frac{d}{dz} \AllInt \left[ \phi_{0T} q_0^2 + \epsilon(2 \phi_{0T} q_0 q_1 +  \phi_{1t} q_0^2) \right] dT =  \epsilon 2 \Real \AllInt i \gamma q_{0TT} q_{0T} dT
\end{equation}
which in turn reduces in the same way as the conservation of energy to 
\begin{equation}
 \phi_{1t}^+  + \phi_{1t}^- = 0
 \label{BlackPhaseBalance}
\end{equation}
\end{subequations}
%%%%%%%%%%%%%%%%%%EXPAND%%%%%%%%%%%%%%%%
By substituting in the asymptotic approximations \eqref{FirstPAsym} found early for $\phi_{1t}^{\pm}$, we arrive at an expression for $t_0$
\begin{equation}
t_{0Z} = 0 
\end{equation}

\begin{figure}
	\begin{center}
	\includegraphics[width= 5in]{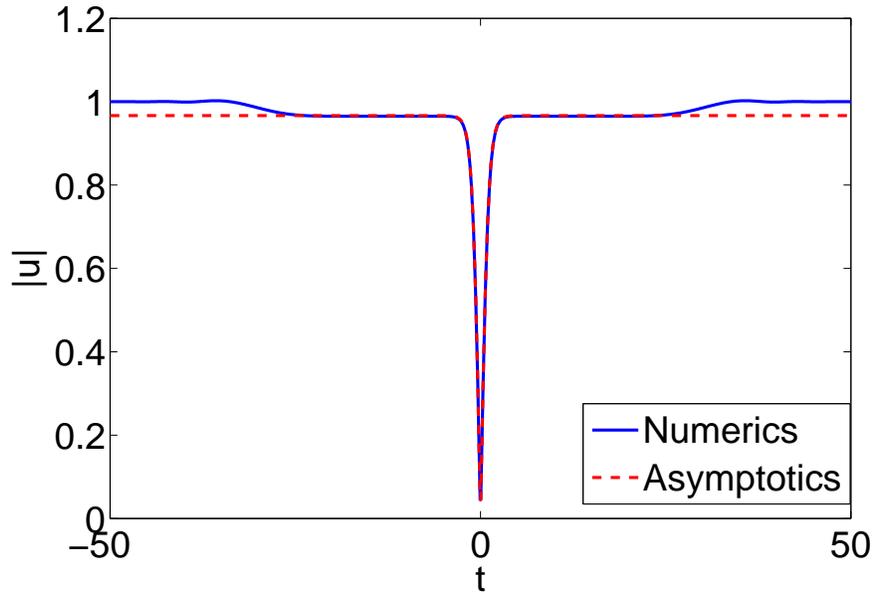}
	\caption{Numerical results plotted against the asymptotic approximation for the magnitude $|u|$ up to $O(\epsilon)$.  Here $z= 30$ and $\epsilon \gamma= 0.05$. }
	\label{Fig: BlackMag}
	
		\end{center}
\end{figure}

\begin{figure}
	\begin{center}

	\includegraphics[width= 5in]{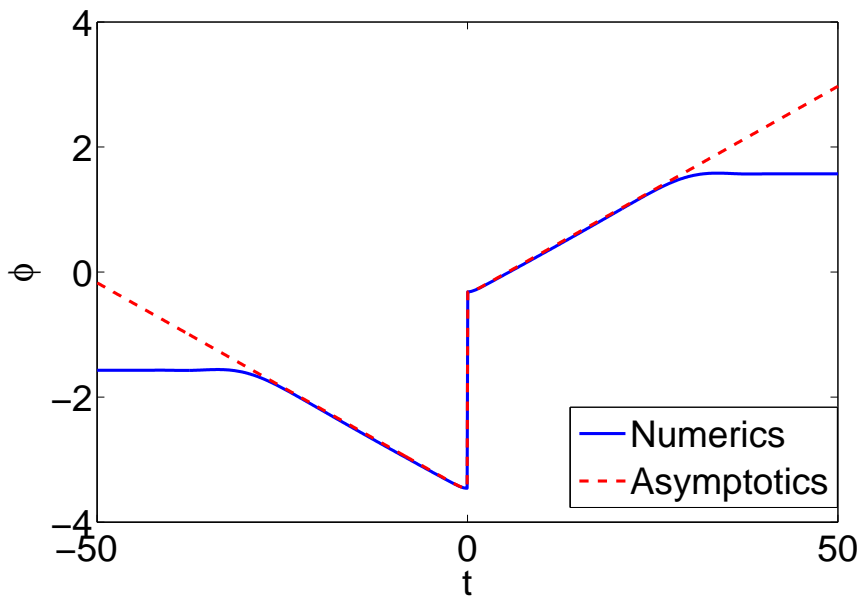}
	\caption{Numerical results plotted against the asymptotic approximation for the phase $\phi$ up to $O(\epsilon)$.  Here $z= 30$and $\epsilon \gamma= 0.05$. }
	\label{Fig: BlackPhase}
	
			\end{center}
\end{figure}

\begin{figure}
	\begin{center}
	
	\includegraphics[width= 5in]{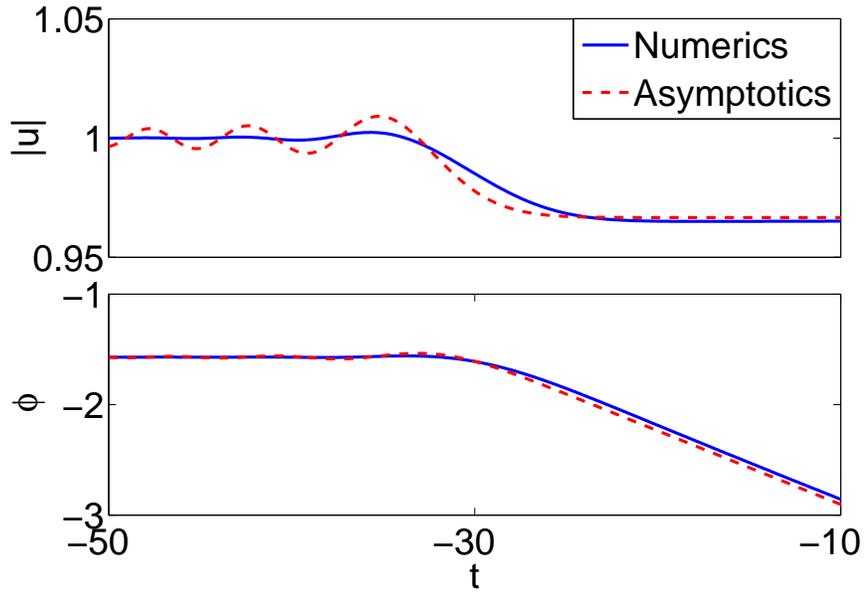}
	\caption{Asymptotic approximation $|u| \approx \uinf + \epsilon w$ and $\phi \approx -\frac{\Delta \phi_{\infty}}{2t} + \epsilon \theta$ for the boundary layer compared to numerics.  Here $z= 30$and $\epsilon \gamma= 0.05$.}
	\label{Fig: ValBorder}
	
			\end{center}
\end{figure}

\begin{figure}
	\begin{center}
	
	\includegraphics[width= 5in]{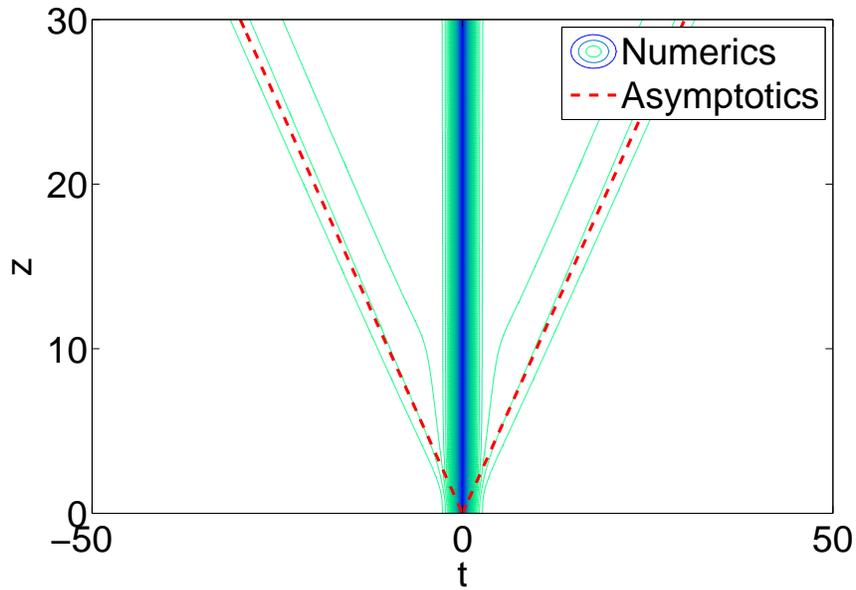}
	\caption{The predicted location of the boundary layer $t_{BL} = \pm \uinf z$ displayed over a contour plot of $|u|$.  Here $\epsilon \gamma = 0.05$. }
	\label{Fig: ConBorder}

	\end{center}
\end{figure}

This can now be compared with numerics.  The magnitude and phase are depicted in Figure \ref{Fig: BlackMag} and Figure \ref{Fig: BlackPhase} respectively.  Here we see the inner region, discussed earlier, is $t \in (-30, 30)$ where the asymptotic solution matches the numerics and rest of the domain constitutes the outer region where the asymptotic solution diverges.  The boundary layer shown in Figure \ref{Fig: ValBorder} compares the solutions \eqref{BLMag} and \eqref{BLPhaseleft} to numerics and illustrates how the inner and outer solutions are connected.  The propagation of this boundary layer can be seen in Figure \ref{Fig: ConBorder} where contour plot contains the soliton extending down the middle and shelf extending out from it.  The speed of the boundary layer matches the speed predicted by our long wave approximation in section \ref{Boundary Layer}.


\clearpage

\section{The Grey Soliton}
\label{Sec: General}

%In our example problem we applied the method of multiple scales, found the solution to the $O(\epsilon)$ equation, and then determined the remaining parameters by requiring the shelf emanating from the soliton satisfy perturbed conservation laws derived from equation \eqref{NormalNLS}.  We apply this method to the more general problem of a grey soliton under the influence of a general perturbation which may not vanish as $t\rightarrow \pm \infty$, but we will see there are some important differences.  For instance, we relax the restriction that $u_0$ satisfy the phase component of the boundary condition.  As we will verify, the over all phase change across both the soliton and the shelf will remain constant and satisfy the boundary condition.  We still require that
%%
% \begin{equation}
% A^2 + B^2 = \uinf^2
% \label{ABU}
% \end{equation}
% %
 
 %
Consider a grey soliton with velocity $A(Z)$ and $ A^2 + B^2 = \uinf^2$.  Let $u = qe^{i\phi}$ where $q>0$ and $\phi$ are real functions of $z$ and $t$ and introduce moving frame of reference $T = t - \int_0^z A(\epsilon s) ds - t_0$ and $\zeta = z$, so equation \eqref{NormalNLS} becomes
\begin{equation}
iu_\zeta  - iAu_T - \frac{1}{2}u_{TT} + (|u|^2 - \uinf^2)u = \epsilon F[u]
\end{equation}
And, then using $u=qe^{i \phi}$
\begin{equation}
i(q_\zeta  + i \phi_\zeta  q) - iA(q_T + i\phi_Tq) - \frac{1}{2} \left[ q_{TT} + i2\phi_Tq_T + (i\phi_{TT} - \phi_T^2)q \right] + q^3 - \uinf^2q = \epsilon F[q,\phi]
\end{equation}
This is now broken into real and imaginary parts
\begin{subequations}
\begin{eqnarray}
q_\zeta  &=& Aq_T + \frac{1}{2} (2\phi_T q_T + q \phi_{TT}) + \epsilon \Imag\left[F[u] \right] \\
\phi_\zeta  q &=&A\phi_Tq -\frac{1}{2}(q_{TT} - \phi_T^2 q) + (|q|^2 - \uinf^2) q + \epsilon \Real\left[F[u] \right] 
\label{MovingRefqphi}
\end{eqnarray}
\end{subequations}
We now write equations \eqref{MovingRefqphi} in terms of the slow evolution variable $\zeta = \epsilon Z $ and series expansions $q  = q_0 + \epsilon q_1 + O(\epsilon^2)$ and $\phi = \phi_0 +\epsilon \phi_1 + O(\epsilon^2)$.  At $O(1)$ the equations are satisfied by the soliton solution \eqref{Fullqphi}.   

At $O(\epsilon)$ we have 
\begin{eqnarray*}
q_{1\zeta } &=&  Aq_{1T} + \frac{1}{2} \left[ 2(\phi_{0T} q_{1T} + \phi_{1T}q_{0T}) + \phi_{1TT} q_0 + \phi_{0TT} q_1 \right] + \Imag \left[ F[u_0] \right]  - q_{0Z}\\
\phi_{1\zeta} q_0 &=& -\phi_{0\zeta} q_1 + A(\phi_{0T} q_1 + \phi_{1T} ) - \frac{1}{2}(q_{1TT} - \phi_{0T}^2q_1 - 2 \phi_{0T} q_0 \phi_{1T}) + 3q_0^2 q_1 - \uinf^2 q_1 \\
&&- \Real \left[ F[u_0] \right] - \phi_{0Z}q_0 
\end{eqnarray*}
We look for stationary solutions at $O(\epsilon)$
\begin{subequations}
\begin{eqnarray}
\label{GreyCorrection1}
0&=&  Aq_{1T} + \frac{1}{2} \left[ 2(\phi_{0T} q_{1T} + \phi_{1T}q_{0T}) + \phi_{1TT} q_0 + \phi_{0TT} q_1 \right] + \Imag \left[ F[u_0] \right]  - q_{0Z} \\
\label{GreyCorrection2}
0 &=& A(\phi_{0T} q_1 + \phi_{1T} q_0) - \frac{1}{2}(q_{1TT} - \phi_{0T}^2q_1 - 2 \phi_{0T} q_0 \phi_{1T}) + 3q_0^2 q_1 - \uinf^2 q_1 \\
&&-  \Real \left[ F[u_0] \right] - \phi_{0Z}q_0 \notag
\end{eqnarray}
\end{subequations}
Where 
\begin{subequations}
\begin{eqnarray}
q_{0Z} &=& \frac{1}{2}\left(AA_Z + BB_Z\tanh^2(x)\right)q_0^{-1} + q_{0T} \left(\frac{B_Z}{B} - t_{0Z} \right) \\
\phi_{0Z} &=& \left(AB_Z - BA_Z\right)\tanh(x)q_0^{-2} + \phi_{0T} \left(\frac{B_Z}{B} - t_{0Z} \right)  + \sigma_{0Z}
\end{eqnarray}
\end{subequations}

\begin{figure}
	\begin{center}
	\includegraphics[width= 5in]{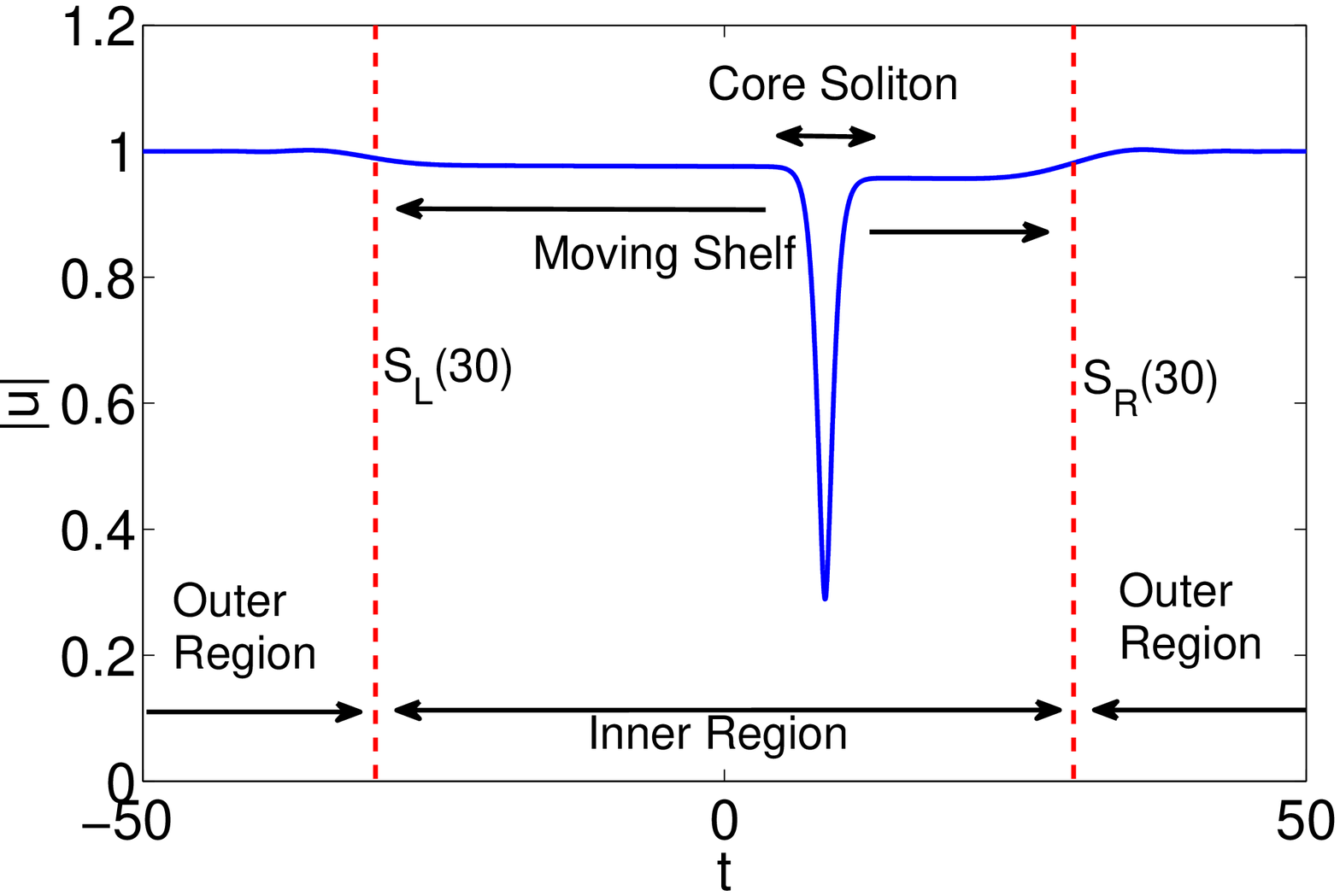}
	  \includegraphics[width= 5in]{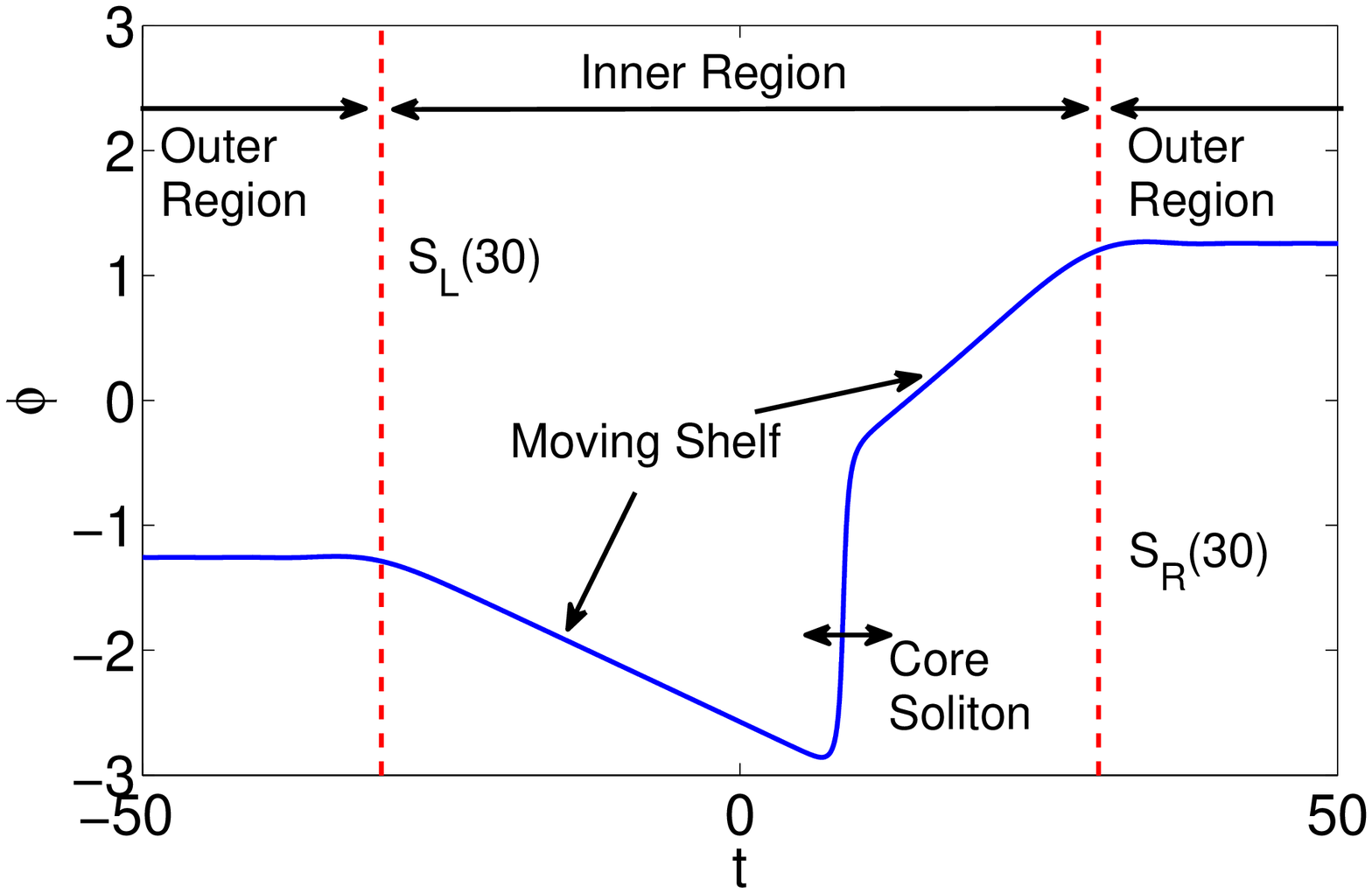}
	\caption{Diagram of a perturbed soliton for both the Magnitude and Phase.  Here $F[u] = i\gamma u_{tt}$, $z = 30$, $\epsilon\gamma = .05$ and $\Delta \phi_0 = 4\pi/5$. }
	\label{Fig: Anatomy}
	
		\end{center}
\end{figure}

Unlike the black soliton problem, we do not solve this explicitly.  We  assume a shelf structure similar to the one found in our example problem will develop; this is supported by numerical computations.  Consider equation \eqref{GreyCorrection1} in the limit $T \rightarrow \pm \infty$ using $q_0 \rightarrow u_{\infty}$ and $u_{\infty Z}= ImF[u_{\infty}]$ yields
\begin{equation}
0 = Aq_{1T}^{\pm} + \frac{\uinf}{2} \phi_{1TT}^{\pm}
\end{equation}
We assume $q_1$ tends to a constant with respect to $t$; i.e. $q_{1T} \rightarrow 0$ as $t \rightarrow \pm \infty$.  As a result $\phi_{1TT} \rightarrow 0 $.  Then $q_1$ and $\phi_{1T}$ both go to constants as $t \rightarrow \pm \infty$ which corresponds to a shelf developing around the soliton.  Now, from equation \eqref{GreyCorrection2} in the limit $T \rightarrow \pm \infty$ we get
\begin{equation}
A \phi_{1T}^{\pm} + 2 \uinf q_1^{\pm} =-\Real\left[ F[\uinf]\right]/\uinf  \pm \displaystyle\frac{\left(AB_Z - BA_Z\right)}{\uinf^2}+ \sigma_{0Z}
\label{EqSet1a}
\end{equation}
We define $\Delta\phi_0$ by
\begin{subequations}
\begin{equation}
\label{phi0}
\Delta \phi_0 = 2\tan^{-1} \left( \frac{B}{A} \right)
\end{equation}
the phase change across the core soliton.  his is consistent with the soliton parameters $A$ and $B$ being expressed in terms of background magnitude, $\uinf$, and phase change, $\Delta \phi_0$,
\begin{equation}
A = \uinf \cos \left( \frac{ \Delta \phi_0}{2} \right)~~~~~~~~~B = \uinf \sin \left( \frac{ \Delta \phi_0}{2} \right)
\label{AandB}
\end{equation}
\end{subequations}
Using equation \eqref{Background2} and substituting in \eqref{AandB} on the right hand side equation \eqref{EqSet1a} becomes
\begin{equation}
A \phi_{1T}^{\pm} + 2 \uinf q_1^{\pm} =\phi_Z^{\pm} \pm \frac{ \Delta\phi_{0Z}}{2} + \sigma_{0Z}
\label{EqSet1}
\end{equation}
We recognize the right hand side as the derivative of the phase on either edge  of the core soliton (See Fig \ref{Fig: Anatomy}).

\section{Grey Conservation Laws}
\label{Sec: GenCon}

Next we use the evolution equations \eqref{ConEvol} to solve for the shelf parameters $q_1^{\pm}$ and $\phi_{1t}^{\pm}$ as well as the slow evolution variables $A$, $\sigma_{0Z}$ and $t_0$.  Note that if we find $A$, then $B = (\uinf^2 - B^2)^{1/2}$.  The edge of the shelf still propagates with velocity $V(Z) = \uinf(Z)$, however the speed may now vary in $z$.  In terms of the moving frame of reference the boundaries of the shelf are
\begin{subequations}
\begin{eqnarray}
S_L(\zeta) &=&  -\int_0^\zeta \left[ \uinf(\epsilon s) + A(\epsilon s) \right]ds \\
S_R(\zeta) &=& ~~\int_0^\zeta \left[ \uinf(\epsilon s) - A (\epsilon s) \right] ds
\end{eqnarray}
\end{subequations}
where $S_L$ and $S_R$ give the position in $T$ of the left and right boundaries of the shelf respectively at $\zeta$.  Note that $A \leq \uinf$ for all $Z$, thus the soliton can not over take the shelf.

We begin with the evolution equation for the Hamiltonian \eqref{dHdz}. 
\begin{equation}
\frac{d}{d\zeta} \AllInt \left[ \frac{1}{2} |u_t|^2 + \frac{1}{2}(\uinf^2- |u|^2 )^2 \right] dt = \epsilon \left(\uinf^2\right)_Z\AllInt \left[\uinf^2 - |u|^2 \right] dt +   2\epsilon \Real \AllInt F[u] u_{\zeta}^* dt 
\end{equation}
Substituting in $u= (q_0 + \epsilon q_1)e^{i(\phi_0 + \epsilon \phi_1)}$ and changing variables to the moving frame of reference, we have up to $O(\epsilon)$
\begin{equation}
\frac{d}{d\zeta} \AllInt (q_{0T}^2 + \phi_{0T}^2q_0^2)+ (\uinf^2 - q_0^2)^2dT = 2 \epsilon \left(\uinf^2\right)_Z \AllInt \left[ \uinf^2 - q_0^2 \right] dt - 4 \epsilon \Real \AllInt F[u_0] A u_{0T}^* dT
\end{equation}
where both here and later on $u_0 = q_0 e^{i\phi_0}$.  The Hamiltonian is unique among the evolution equations \eqref{ConEvol} in that the contribution of the shelf appears only at $O(\epsilon^2)$ or higher and may be ignored.  We now put in the soliton form \eqref{Fullqphi} to get 
\begin{equation} 
2B^2B_Z = (\uinf^2)_ZB -  A\Real \AllInt F[u_0] u_{0T}^* dT
\label{Hstep2}
\end{equation}
Taking a derivative with respect to $Z$ of the equation $\uinf^2 = A^2 + B^2$ we get
\begin{equation}
(\uinf^2)_Z = 2 A A_Z + 2 B B_Z
\end{equation}
which can be used consolidate equations \eqref{Hstep2} down to
\begin{equation}
2BA_Z = \Real \AllInt F[u_0] u_{0T}^* dT
\label{EqSet12},
\end{equation}

The evolution equations for energy \eqref{ConEnergy} and momentum \eqref{ConMo} both remain the same after transforming to the moving frame of reference 
\begin{eqnarray}
\label{TConEnergy}
 \frac{d}{d\zeta }\AllInt \left[\uinf^2 - |u|^2 \right]  dT& = &2 \epsilon  \Imag \AllInt\left[ F[\uinf] \uinf - F[u] u^*\right] dT\\
 \frac{d}{d\zeta } \Imag \AllInt u u_T^* dT &= &2 \epsilon \Real \AllInt F[u] u_T^*dT
\end{eqnarray}
The inner region over which $q_1$ and $\phi_1$ are relevant is $T \in [S_L(\zeta)  , S_R(\zeta) ]$, and outside this region $q1 = \phi_{1T} = 0$.  At $O(1)$ the equations are satisfied and at $O(\epsilon)$ we have
\begin{subequations}
\begin{eqnarray}
B_Z -  \frac{d}{d \zeta} \MInt q_0 q_1dT &=& \Imag \AllInt \left[ F[\uinf] \uinf - F[u_0] u_0^*\right] dT \\
-2(AB)_Z - \frac{d}{d \zeta} \MInt \left[ 2 \phi_{0T} q_0 q_1 +  \phi_{1T} q_0^2 \right] dT &=& 2  \Real \AllInt F[u_0] u_{0T}^*dT
\end{eqnarray}
\end{subequations}
Since the integrands on the left hand side are not functions of $\zeta$, we can apply the fundamental theorem of calculus to arrive at 
\begin{subequations}
\begin{eqnarray}
\label{EqSet2}
B_Z - \uinf \left[(\uinf - A) q_1^+ + (\uinf +A) q_1^- \right] &=& \Imag \AllInt \left[ F[\uinf] \uinf - F[u_0] u_0^*\right] dT \\
 \label{EqSet3}
2(AB)_Z + \uinf^2 \left[ (\uinf -A) \phi_{1T}^+ +  (\uinf+A) \phi_{1T}^- \right] &=& -2 \Real \AllInt F[u_0]u_{0T}^*dT
\end{eqnarray}
\label{EandI}
\end{subequations}
We are left now with the evolution of the center of energy
\begin{equation}
\frac{d}{d\zeta } \AllInt t ( \uinf^2 - |u|^2) dt = -\Imag \AllInt u u_t^*  dt +
 2\epsilon \Imag \AllInt t \left( F[\uinf] \uinf -  F[u] u^* \right) dt 
\end{equation}
which after transforming to the moving frame of reference is now
\begin{eqnarray*}
\frac{d}{d\zeta } \AllInt  \textrm{\Large(} T + \int_0^{\zeta} A+ t_0 \textrm{\Large)}  (\uinf^2 - |u|^2) dT &=& -\Imag \AllInt u u_T^* dT \\
&+& 2 \epsilon \Imag \AllInt\textrm{\Large(} T + \int_0^{\zeta} A+ t_0 \textrm{\Large)}  \left( F[\uinf] \uinf - F[u] u^* \right) dT
\end{eqnarray*}
After rearranging some terms we have
\begin{subequations}
\begin{eqnarray}
\label{CCE0}
&\displaystyle\frac{d}{d\zeta } &\AllInt T( \uinf^2 - |u|^2 ) dT\\
\label{CCE1}
 &&+ \left( \int_0^{\zeta} A + t_0 \right) \left[ \displaystyle\frac{d}{d\zeta } \AllInt \left[ \uinf^2 - |u|^2 \right] dT - \epsilon 2 \Imag \AllInt \left( F[\uinf] \uinf - F[u] u^* \right) dT\right] \\
 \label{CCE2}
&&+ A    \AllInt \left[ \uinf^2 - |u|^2 \right]  dT  + \Imag \AllInt u u_T^* dT\\
\label{CCE3}
&&= -\epsilon t_{0Z} \AllInt \left[\uinf^2 - |u|^2 \right] dT + 2 \epsilon \Imag \AllInt T  \left( F[\uinf] \uinf - F[u] u^* \right)dT 
\end{eqnarray}
\label{CCE}
\end{subequations}
Line \eqref{CCE0} yields
\begin{equation}
\frac{d}{d\zeta } \AllInt T(\uinf^2 - |u|^2 ) dT = -2  \left[ S_R (\uinf - A)  q_1^+ + S_L (\uinf + A)  q_1^- \right] \uinf
\end{equation}
 The terms on line \eqref{CCE1} are the energy equation \eqref{TConEnergy} and cancel out.  The terms on line \eqref{CCE2} are calculated up to $O(\epsilon)$ using the previous results \eqref{EandI}
 \begin{eqnarray*}
 E(\zeta ) &=& 2B - 2   \left[S_R(Z) q_1^+ - S_L(Z) q_1^- \right] \uinf + \epsilon E_1(Z) +O(\epsilon^2) \\
 I(\zeta ) &=& -2AB -  \uinf^2 \left[ S_R(Z) \phi_{1t}^+ - S_R(Z) \phi_{1t}^-\right] +\epsilon I_1(Z) +O(\epsilon^2) 
 \end{eqnarray*}
Noting that $S_R$ and $S_L$ are $O(1/\epsilon)$ in terms of $Z$.

When we put everything together in terms of slow evolution variable $Z = \epsilon \zeta$ we get from \eqref{CCE}
\begin{eqnarray}
&& \epsilon 2 B t_{0Z}=2 \epsilon  \Imag \AllInt T \left( F[\uinf] \uinf - F[u_0] u_0^* \right)dT \notag \epsilon + AE_1(Z)+ \epsilon I_1(Z)   \\
 &&~~~~~  +\textrm{\LARGE[}  2\uinf \left[ S_R (\uinf - A)   q_1^+ + S_L (\uinf + A)  q_1^- \right] + 2 \uinf A \left[S_R q_1^+ -S_Lq_1^- \right] \\
   &&~~~~~ +  \uinf^2 \left[ S_R \phi_{1t}^+ - S_L \phi_{1t}^-\right] \textrm{\LARGE]}   \notag
\end{eqnarray}
This breaks into $O(1)$ terms
\begin{equation}
\label{EqSet4}
2 \left[ S_R q_1^+ + S_L q_1^- \right] + \left[ S_R \phi_{1T}^+ - S_L \phi_{1T}^- \right] = 0
\end{equation}
and $O(\epsilon)$ terms which include higher order energy and momentum terms have not been determined
The six equations \eqref{EqSet1}, \eqref{EqSet12}, \eqref{EqSet2}, \eqref{EqSet3} and \eqref{EqSet4} can now be used to solve for the full set of seven parameters $q_1^{\pm}$, $\phi_{1t}^{\pm}(= \phi_{1T}^{\pm})$,  $A$, $\sigma_0$, and $t_0$.  

\framebox[7in][l]{
\parbox[b]{6.5in}{
\begin{subequations}
\begin{eqnarray}
\frac{d}{dZ}\uinf &=&\Imag \left[ F[\uinf] \right]\\
2B\frac{d}{dZ}A &=& \left( \Real \AllInt F[u_0] u_{0T}^*dT\right) \\
\uinf\frac{d}{dZ}\sigma_{0} &=& \left(  B_Z - \Imag \AllInt  F[\uinf] \uinf - F[u_0]u_{0}^* dT + \Real [F[\uinf]]\right) \\
q_1^+ &=& \frac{1}{2}\left( \sigma_{0Z} + \Delta \phi_{0Z} \right)/\left(\uinf-A\right) \\
q_1^- &=& \frac{1}{2}\left( \sigma_{0Z} - \Delta \phi_{0Z} \right)/\left(\uinf+A\right) \\
\phi_{1T}^+ &=& -2q_1^+\\
\phi_{1T}^- &=& 2q_1^-\\
&& \notag\\ 
\label{Extra1}
B_Z &=& \left( \uinf u_{\infty Z} - A A_Z \right)/B\\
\label{Extra2}
 \Delta\phi_{0Z} &=& \left( 2AB_Z - 2BA_Z\right)/\uinf^2
\end{eqnarray}
\label{Everything}
\end{subequations}

}}
These equation may now be easily solved from top to bottom.  We have added equations \eqref{Extra1} and \eqref{Extra2} to the list since it is often better to use these formulation for $B_Z$ and $\Delta\phi_{0Z}$ rather then working out $B$ and $\Delta\phi_{0}$ explicitly and then taking derivatives.

The equations found can also be used to show that the soliton and shelf taken together do not break the phase component of the boundary condition.  By combining equations \eqref{EqSet12} and \eqref{EqSet3} we arrive at
\begin{equation}
2(AB)_Z +  \uinf^2  \left[ (\uinf -A) \phi_{1T}^+ + (\uinf + A) \phi_{1T}^- \right] = 4BA_Z
\end{equation}
which may be rewritten as
\begin{equation}
2AB_Z - 2BA_Z + \uinf^2 \frac{d}{d\zeta} \left[  \phi_1(S_R) - \phi_1(S_L) \right] = 0
\label{ZeroPhase1}
\end{equation}
If we define $\phi_1$ as follows
\begin{equation}
\Delta \phi_1 = \phi_1(S_R) - \phi_1(S_L)
\end{equation}
then $\epsilon \phi_1$ is the phase change across the shelf. Substituting this definition along with \eqref{phi0} into equation \eqref{ZeroPhase1} we arrive at
\begin{equation}
\frac{d}{dZ} \Delta \phi_0  + \epsilon \frac{d}{dZ} \Delta \phi_1 = 0 
\end{equation}
Thus, the total phase change across the inner region remains constant, which agrees with our earlier result that $\Delta \phi_{\infty}$ (the phase change from $-\infty$ to $\infty$) remains constant for all perturbations.

\pagebreak

\section{$t_{0Z}$ and Asymptotic Behavior}

To find the final parameter $t_0$ we must find a first order solution.  We look for series solution to equation \eqref{NLSPert} of the form $u = u_0 + \epsilon u_1 + O(\epsilon^2)$, and at $O(\epsilon)$ we have
\begin{equation}
iu_{1z}  + \left(- \frac{1}{2} \partial_t^2 + 2|u_0|^2 - \uinf^2 \right)u_1   + \left(u_0^2 \right)\bar{u_1} =  F[u_0] - i u_{0Z}
\end{equation}
or after changing variables to moving frame of reference $T = t - \int_0^\zeta A(\epsilon s) ds - t_0$, $z= \zeta$
\begin{equation}
i u_{1\zeta}  + \left(-iA\partial_{T} - \frac{1}{2} \partial_{T}^2 + 2|u_0|^2 - \uinf^2 \right)u_1   + \left(u_0^2 \right)u_1^* = F[u_0] - i u_{0Z}
\end{equation}
Here
\begin{equation}
u_{0Z} = A_Ze^{i\sigma} + \frac{B_Z}{B} \left(  u_0 - A e^{i \sigma}  \right) + u_{0 T}\left( -t_{0Z} +  \frac{B_Z}{B} T \right) + i\sigma_z u_0
\label{orderE}
\end{equation}

If we look for stationary solutions, this can be written as a system of coupled second order differential equations
\begin{subequations}
\begin{equation}
L \mathbf{U}_1 = \mathbf{G}[u_0]
\end{equation}
where
\begin{equation}
\mathbf{U}_1 = \left( \begin{array}{c} \Real [u_1]\\ \Imag[u_1] \end{array} \right)~~~~~~~~~~
\mathbf{G}[u_0]  = \left( \begin{array}{c}  \Real\left[F[u_0] - i u_{0Z}\right] \\  \Imag\left[F[u_0] - i u_{0Z}\right] \end{array} \right)
\end{equation}
and
\begin{equation}
L = \left[ \begin{array}{cc} -\frac{1}{2} \partial_T^2 + \left(3A^2+ B^2\tanh(BT) - \uinf^2 \right)&  A\partial_T + 2AB\tanh(BT)\\
                                                -A\partial_T + 2AB\tanh(BT)    &  -\frac{1}{2} \partial_T^2 + \left(A^2+ 3B^2\tanh(BT) - \uinf^2 \right)
                                               \end{array} \right]
\end{equation}
\label{LinNLS}
\end{subequations}

This system has homogeneous solutions
\begin{subequations}
\begin{eqnarray}
\mathbf{U}_{11}& =&\left( \begin{array}{c} 0\\ \sech^2(BT)\end{array} \right)\\
\mathbf{U}_{12} &=& \left( \begin{array}{c} B\tanh{BT} \\-A\end{array} \right)\\
\mathbf{U}_{13}& =& \left( \begin{array}{c} B(BT\tanh(BT) - 1) \\ A\left(- B T +  \frac{3}{2}  BT \sech^2(BT) + \frac{3}{2}  \tanh(BT)\right)\end{array} \right)\\
\mathbf{U}_{14}&=&\left( \begin{array}{c} - \frac{4AB}{A^2 - B^2}\cosh^2(BT) \\ 3BT\sech^2(BT) + 4\tanh(BT) + \tanh(BT)\cosh(2BT)\end{array} \right)
\end{eqnarray}
\end{subequations}
and using the method of undetermined coefficients we can obtain a particular solution.

To put $u_1$ in terms of our magnitude and phase functions $q_0$, $q_1$, $\phi_0$ and $\phi_1$, we expand our previous approximation for $u$
\begin{equation}
u = (q_0+\epsilon q_1)e^{i(\phi_0 + \epsilon\phi_1)} =   q_0e^{i\phi_0} + \epsilon \left(q_1 + i\phi_1q_0\right)e^{i\phi_0} +  O(\epsilon^2)
\end{equation}
so that
\begin{eqnarray}
u_0 &=&  q_0e^{i\phi_0} \\
u_1 &=&  \left(q_1 + i\phi_1q_0\right)e^{i\phi_0} \\
&=&\left[ q_1\cos(\phi_0) - \phi_1q_0\sin(\phi_0) \right] + i \left[q_1\sin(\phi_0) + \phi_1q_0\cos(\phi_0) \right]
\end{eqnarray}

Finally, $t_{0Z}$ is determined be taking the asymptotic behavior of the solution $u_1$ as $t\rightarrow \pm\infty$
\begin{equation}
u_{1t}^{\pm} = -\phi_{1t}^{\pm}(\pm B) + i \phi_{1t}^{\pm}(A)
\end{equation}

\section{Example for Grey Solitons}
\label{Sec: Dissipation}

\begin{figure}
	\begin{center}
	\includegraphics[width= 5in]{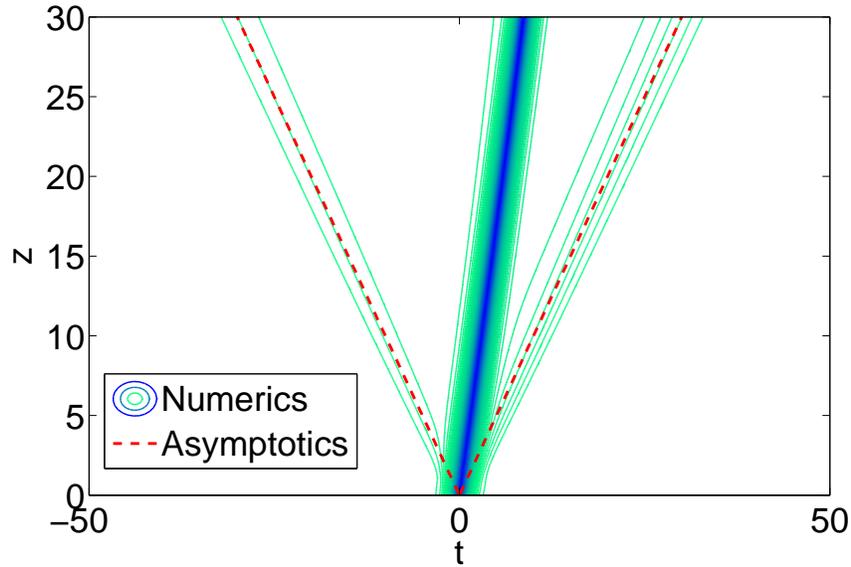}
	\caption{The predicted shelf edge overlaid on the contour plot of numerical results.  Here $\epsilon \gamma= .05$ , and $\Delta \phi_0 = 4 \pi/5 $. }
	\label{Fig: GreyConBorder}
	
		\end{center}
\end{figure}

Let us return to the perturbation $F[u] = i\gamma u_{tt}$, however we now consider the evolution of a general dark soliton with $\uinf(0) = 1$.   As was the case for black solitons, the background height $\uinf$ is found to be constant from equations \eqref{Background2}.  In Fig \ref{Fig: GreyConBorder} we see that the velocity of the soliton does not effect the velocity of shelf which still moves with velocity $V = \pm \uinf$.  Using the equations derived in sections \ref{Sec: General} and \ref{Sec: GenCon} we now solve for all relevant parameters
\begin{subequations}
\begin{eqnarray}
A_Z &=& ~~0\\
\sigma_{0Z} &=&- \frac{4}{3} \gamma \uinf^2 \sin^3 \left( \alpha \right)\\
q_1^+ &=& - \frac{2}{3} \gamma\left(   \uinf +A \right) \sin\left( \alpha \right)\\
q_1^- &=& - \frac{2}{3} \gamma \left(  \uinf -A  \right) \sin\left( \alpha \right) \\
\phi_{1t}^+ &=&~~\frac{4}{3} \gamma\left(   \uinf +A \right) \sin\left( \alpha \right) \\ 
\phi_{1t}^-  &=&  -\frac{4}{3} \gamma \left(  \uinf -A  \right) \sin\left( \alpha \right) 
\end{eqnarray}
\end{subequations}
where $\alpha = \frac{\Delta \phi_0}{2}$ and $\Delta \phi_0$ is the phase change across the core soliton as defined in equation \eqref{phi0}.

\begin{figure}
	\begin{center}
	\includegraphics[width= 5in]{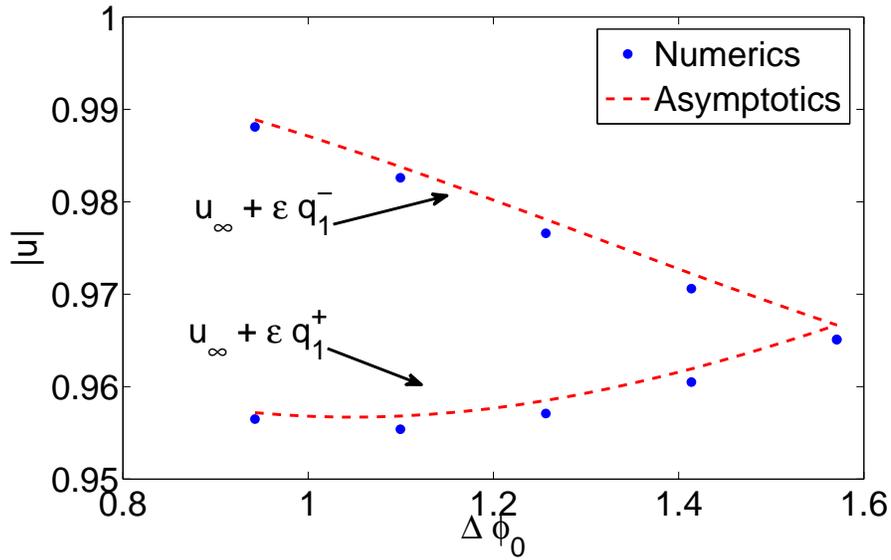}
	\caption{The shelf height found numerical for various values of $\Delta\phi_0$ plotted over the asymptotic approximations.  Here $\uinf = 1$ and $\epsilon \gamma= .05$. }
	\label{Fig: GreyShelves}
	
		\end{center}
\end{figure}

\begin{figure}
	\begin{center}
	\includegraphics[width= 5in]{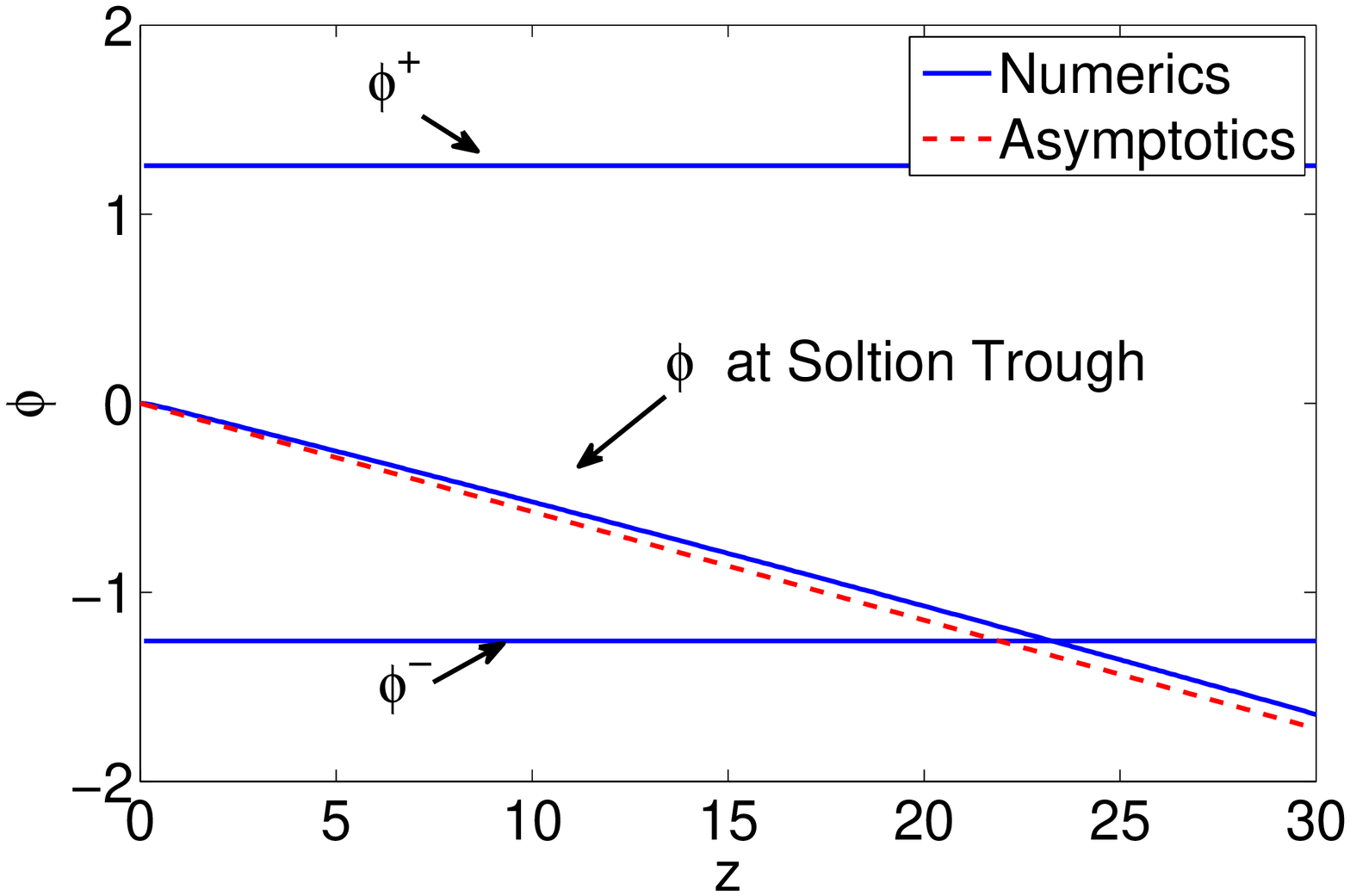}
	\caption{$\sigma_0(Z)$ plotted against the phase at plus and minus infinity along with the phase at the center of the soliton.  Here $\uinf= 1$, $\epsilon \gamma= .05$ , and $\Delta \phi_0 = 4 \pi/5 $. }
	\label{Fig: GreyPhaseDiff}
	
		\end{center}
\end{figure}

Unlike the speed of the shelf, the magnitude of the shelf does depend on the soliton's velocity (which is in turn related to the soliton's depth and the phase across the soliton).  As illustrated in Fig \ref{Fig: GreyShelves} the shelf grows shallower behind the soliton for lager speeds (or smaller phase change $\Delta \phi_0$).  The extra phase $\sigma_0(z) =-\epsilon z  \frac{4}{3} \gamma \uinf^2 \sin^3 \left( \alpha \right)$ induced by the perturbation means that the spatial frequency of the soliton is different then the frequency of the cw background that it lies on.  Though $\sigma_{0}$ evolves adiabatically the soliton eventually becomes noticeably out of phase from the background as shown in Fig \ref{Fig: GreyPhaseDiff}.  Here the background phase ($\phi^+$ and $\phi^-$) is constant since the fast evolution of the background phase was take out in equation \eqref{NLSPert}.

\section{Conclusion}

In conclusion, we have derived a new approach to dark soliton perturbation theory which break the problem into an inner region around the soliton and an outer region equal to the boundary at infinity.  We find that under perturbation a dark soliton develops a shelf the edge of which propagates out at speed equal to the magnitude of the cw background.  Analytically this shelf arises due to the difference between the perturbed soliton and the soliton solution of unperturbed NLS which satisfies the boundary conditions.  It was shown to be possible for the soliton to have a different frequency then the cw background.  The method extends to general perturbations and works for both moving and constant background.  For example perturbations the asymptotic approximation calculated was compared to numerical results.  These comparisons confirm the existence of the analytically predicted shelf and supported our claim that the adiabatic approach alone is insufficient to fully describe the dark soliton evolution.  The main conclusion is that the non-vanishing background and soliton must be treated separately to have a consistent perturbation theory for dark solitons.

\appendix

\bibliography{MainBib}

\end{document}